\documentclass[aps,preprint,groupedaddress]{revtex4}
\usepackage{graphics}
\usepackage{bm}

\usepackage{amsmath}    
\usepackage{amsfonts}
\usepackage{amssymb}
\usepackage{graphicx}   
\usepackage{amssymb}
\usepackage{amsfonts}

\begin{document}

\title{ A Random Matrix model of black
 holes}
\author{Da Xu}

 \affiliation{the University of Iowa, Department of Mathematics, Iowa City,
IA, 52246,USA}
 \email{dxu@math.uiowa.edu}   
\date{\today}

\begin{abstract}
A random matrix model of black holes is given based on analysis of Gaussian complex ensembles, based on the generalization of chRMT of QCD. Spacetime freedoms are incorporated in terms of eigenvalues of the ensemble. Classical observables of black holes can be computed using this model and it satisfies asymptotic freedom and holographic principle.

\begin{footnotesize}{\emph{PACS}:04.60.-m;11.10.Jj;11.25.Yb}\end{footnotesize}
\end{abstract}

\maketitle

\section{introduction}
  Random matrices are matrices which entries are random variables.
Random matrices were first introduced by Hsu \cite{Hsu} and then
intensively studied by physicists in nuclear physics\cite{M}. Random
matrix theory(RMT), has been a very successful theory in the
analysis of nuclear spectrum. Then it had been applied to condensed
matter physics and recently in QCD physics,2D string theory and
etc\cite{Ma}\cite{V}. In \cite{B}, the authors conjectured that M
theory is equivalent to $N\rightarrow\infty$ supersymmetric matrix
quantum mechanics describing $D_0$ branes. RMT was concerned with
black hole physics in \cite{Pa} to estimate the information lost in
black hole radiation by computing $S$ matrix. In the present work,
we shall propose a new idea to compute black hole radiation based on
RMT.

\section{the statistical nature of measurement of time}
 It has been recognized that there is a deep relation among
general relativity, thermodynamics and quantum mechanics, since
classical black hole thermodynamics was found at 1970s \cite{H}.
Although, it is manifest that a full quantum treatment of gravity
has to involve thermodynamical arguments, the spacetime geometry and
thermodynamics are dealt separately in unified theories so far.
Since Hawking effect, which is mathematically equivalent to Unruh
effect, reveals the thermodynamical properties of black holes, which
is a new effect at large energy scale, compared to classical general
relativity. Therefore, thermodynamical nature of spacetime or time,
has to be considered in a consistent quantum gravity theory.
\cite{X} is concerned with the fundamental difficulty of quantum
gravity from a philosophical and historical point of view.

In the present paper, we would like to relate the nature of time to
how we measure time. Since 1949, the most accurate measurement of
time is based on the mechanism of radiation of cesium atoms.One
second is defined as the duration of 9,192,631,770 cycles of
radiation corresponding to the transition between two energy levels
of the ground state of the cesium-133 atom. Although, this accuracy
is far away from plank scale, we still conjecture any measurement of
time has to involve statistical behavior of a physical system.
Therefore instead of writing down explicit time coordinate, we
consider the statistical behavior of a gravitational system when we
consider quantum gravity issues. In one word, we believe gravitation
is a probabilistic potential. We shall see this physical idea is
realized by random matrix theory.

\section{spacetime freedom in the RMT model of black holes}
As we know, at high energies, QCD can be well understood by perturbation theory, by the virtue of asymptotic freedom.
The chiral symmetry is restored at about 150 MeV by lattice QCD stimulation \cite{K}.
In \cite{V1}, the authors showed that constraints imposed by chiral symmetry and its spontaneous symmetry breaking
determine the structure of low-energy effective partition for the Dirac spectrum.The exact results for the low-lying eigenvalues of the QCD Dirac operator. The statistical properties of these eigenvalues are universal, which is a the property of chiral random matrix model.  The partition function of chRMT \cite{V2} is

\begin{align}
Z^\beta_{N_f,\nu}(m_1,\cdots,m_{N_f})=\int DW \prod^{N_f}_{f=1}\det(D+m_f) e^{-\frac{N\beta}{4} Tr \ v(W^{+}W)},
\end{align}
where $\beta$ is the Dyson index,
\begin{align}
D=\left(\begin{array}{cc}
0 & i W\\
i W^{+} & 0
\end{array}
\right),
\end{align}
and $W^{+}$ is an $n\times m$ matrix with $\nu=n-m$ and $N=n+m$. $N_f$ is the number of flavors of quarks.

The masses of quarks can be thought of taking complex values to involve the $\theta$ angle which causes CP violation(see \cite{V}). Therefore $D+m_f$ is neither hermitian, nor anti hermitian either.
It is tempt to generalize this partition function to black hole states. At extremely high energy or high temperature,  it is meaningful to  enlarge the chiral symmetry group $\rm SU(N_f)\times SU(N_f)$ to
$\rm U(2N_f)$. Use the same $W$ to denote arbitrary $2N_f\time 2N_f$ complex matrices. The partition function of this black hole is
\begin{align}
Z_B(N)=\int DW \det W e^{- \frac{\beta}{4} Tr \ v(W^{+}W)}.\label{black}
\end{align}
The parameter $N$ can be considered as the dimensionless volume of spacetime; $v$ is the potential function. It is reasonable to assume the spacetime freedoms are incorporated to the complex  eigenvalues of the random matrix $W$, because the Dirac operator
$$
D=\left(\begin{array}{cc}0 & iW\\
                         iW^{+} & 0
                         \end{array}
                         \right)=\gamma^\mu \partial_\mu
$$
invovles partial derivatives on spacetime coordinates.
  This is the essential assumption of the present work.
Integrating out the gauge freedoms, we get the partition function in terms of eigenvalues
\begin{align}
Z_B(N)=\int d\lambda \prod_{i=1}^{N}\lambda_i e^{-\frac{\beta}{4}\sum_{i=1}^N \lambda_i}\prod_{i<j}|\lambda_i-\lambda_j|^2,
\end{align}
where $\lambda_i$s are complex eigenvalues(see chapter 15, \cite{M}).   It has been shown that the microscopic eigenvalue correlation at 0(which is the normalized correlation such that the mean spaceing is unit at the point) is universal. So we set $v=Tr(WW^{+})$.

\section{asymptotic freedom property}
The asymptotic behavior of the coupling constant in strong interaction is one of the most important
features in QCD. In the context of renormalization group\cite{W}, the effective coupling constant of QCD has asymptotic freedom, because there $n_f=6$ flavors which makes
$\beta(g)=-\frac{g^3}{4\pi^2}(11/4-1/6 n_f)+O(g^5)$.
We shall provide a alternative expanation of asymptotic freedom using the present model.
So when the energy $E\rightarrow \infty$, when the gavitational effect has to arise, then expalnation is not satifactory.
But in the present model, spacetime freedoms are the quaternion eigenvalues of the random matrix.  As we know, as  the rank of the random matrix $N$ goes to infinity,the normalized(mean spaceing is normalized to be unit) two point correlation function
\begin{align}
R_{22}(r)=1-(\frac{\sin(\pi r)}{\pi r})^2 \label{two point}
\end{align},
where $r=|\lambda_1-\lambda_2|$.
(see\cite{M}). $R_{22}$ goes to 1 as $r$ goes to infinity, and goes to zero as r goes to zero.
The bulk of this normalized two point correlation function is a constant equal to 1. Since the fermion derterminant in the integrand of the partition function is separable for the eigenvalues. Therefore the eigenvalues in the bulk are statistically independent. This fact corresponds the asymptotic freedom in QCD.   In reality, the rank $N$ is finite.
then the correlation functions vanish when $r$ is large.

\section{Gaussian  ensembles}
\subsection{the $\rm U(n)$ symmetry of Gaussian ensembles}
In the analysis of complicated nuclear spectrum, some statistical
distribution on eigenvalues of GOE(Gaussian Orthogonal Ensemble) are
computed\cite{M}. The eigenvalues corresponds to nuclear energy
levels which are always positive. We shall use Gaussian complex ensembles to represent black holes.

Gaussian complex ensemble  is defined on the space of
$N\times N$ matrices with complex entries by the
following properties:\\
(1) The probability $P(H)dH$ that a system of $E_{R4G}$ will belong
to the volume element
\begin{align}
dH=\prod_{\gamma=0}^2\prod_{1\leq k,j\leq N}dH^\gamma _{kj},
\end{align}
where $H^\gamma_{ij}$,$\gamma=0,1,2,3$ which are the four components
of the quaternion $H_{ij}$, is invariant under every automorphism
\begin{align}
H\rightarrow U^+ H U
\end{align}
on the space of quaternion matrices, where $U$ is any unitary
matrix.\\
(2) Various linearly independent components of $H$ are also
statistically independent.

As we know, an invariant of matrix spaces under similarity
transformation only depends on the first $N$ traces. However, we
shall use a stonger lemma. Smooth invariants of $N\times N$ matrices under
similarity transformation,only depends on all the $N$ eigenvalues
of a matrix. In fact, $N\times N$ complex quaternion matrices is a
$2N^2$ real dimensional manifold. Since nonsingular matrix space is
$N^2$ dimensional, the action of the center of nonsingular matrices
which are diagonal nonsingular matrices, on $N\times N$ complex
matrices is invariant on diagonal complex matrices which is of real
dimension $2N$. So the quotient orbifold is $2N$ real dimensional.
The orbiford is just diagonal matrix modular by the permutation
group. We know for a continuous function defined on a matrix space
which is invariant under similarity transformations must be a
function of all the eigenvalues of the matrix.However, if we only
assume that the similarity transformation restricted as unitary
matrices, we have to require the function is analytic. Without
proving, we state the following lemma.
  \emph{$\rho$ is a analytic function on the space of complex
matrices, which is invariant under unitary similarity
transformation. Then $\rho$ only depends on the eigenvalues of a
complex matrix. }

We generalize Porter and Rosenzweig theorem \cite{Porter}.
\emph{
$P(H)$ is an ensemble on the space of $N\times N$ matrices with complex entries, then $P(H)$ is a Gaussian complex
ensemble, if and only if
\begin{align}
P(H)=exp(-a \ tr(HH^+) +b Re \ tr(H)+c),
\end{align}
where $a,b,c$ are real and $a$ is positive,  $c$ is a normalization
constant}

In fact, since $P$ is invariant under unitary transformations, we
shall consider how $H$ is changed under rotation transformations
which is a subgroup of unitary matrices. Let
\begin{align}
H=U^{+}H^{'}U,
\end{align}
where
\begin{align}
U=\left(
\begin{array}{ccccc}
\cos\theta & \sin\theta &0 &...&0\\
-\sin\theta &\cos\theta &0 &...&0\\
0 & 0 &1 &...&0\\
0&0&0&...& 1
\end{array}
\right).
\end{align}
Differentiating the above equation, we get
\begin{align}
\frac{\partial H}{\partial\theta} & =\frac{\partial
U^+}{\partial\theta}H^{'} U+U^{+}H^{'}\frac{\partial U}{\partial\theta}\\
&=\frac{\partial U^{T}}{\partial\theta}UH+HU^{T}\frac{\partial
U}{\partial\theta}.
\end{align}
And
\begin{align}
\frac{\partial U^{+}}{\partial\theta}U=(U^{T}\frac{\partial
U}{\partial\theta})^{T}=
 \left(
\begin{array}{cccc}
 0&-1& 0&0\\
1&0&0&0\\
...&...&...&...\\
0&0&0&0\\
\end{array}
\right)
\end{align}
Since each entry is independent random variables,
\begin{align}
P(H)=\prod_\gamma \prod_{1\leq i,j\leq N}f^\gamma_ij(H^\gamma_{ij}).
\end{align}
 By the definition of Gaussian complex ensemble, $P(H)$ is
invariant under the transform of $U$, which yields
\begin{align}
\frac{\partial \ln
P(H)}{\partial\theta}=\sum_{\gamma,i,j}\frac{\partial\ln
f^{\gamma}_{ij}}{\partial H^{\gamma}_{ij}}\frac{\partial
H^{\gamma}_{ij}}{\partial\theta}=0.
\end{align}
If we write it explicitly, we have the following equation
\begin{align}
\sum_\gamma ( -\frac{\partial\ln f^\gamma_{11}}{\partial
H^\gamma_{11}}+\frac{\partial\ln f^\gamma_{22}}{\partial
H^\gamma_{22}})(H^\gamma_{12}+H^\gamma_{21})+
\sum_\gamma(\frac{\partial \ln f^\gamma_{12}}{\partial
H^\gamma_{12}}+ \frac{\partial
\ln f^\gamma_{21}}{\partial H^\gamma_{21}})(H^\gamma_{11}-H^\gamma_{22})\nonumber\\
+\sum_\gamma \sum_{k\geq 3}(-\frac{\partial\ln
f^\gamma_{1k}}{\partial
H^\gamma_{1k}}H^\gamma_{2k}+\frac{\partial\ln
f^\gamma_{2k}}{\partial H^\gamma_{2k}}H^\gamma_{1k}) +\sum_\gamma
\sum_{k\geq 3}(-\frac{\partial\ln f^\gamma_{k1}}{\partial
H^\gamma_{k1}}H^\gamma_{k2}+\frac{\partial\ln
f^\gamma_{k2}}{\partial H^\gamma_{k2}}H^\gamma_{k1})
\end{align}

These for sum terms depend on different variables. So each of them
is a constant and the sum of the constants is zero. For each
$\gamma$, we have an equation

\begin{align}
-\frac{\partial\ln f^\gamma_{1k}}{\partial
H^\gamma_{1k}}H^\gamma_{2k}+\frac{\partial\ln
f^\gamma_{2k}}{\partial H^\gamma_{2k}}H^\gamma_{1k}=C,
\end{align}
Where $C$ is a constant. $C$ must be zero because the first term of
(2.11) is a product of $H^\gamma_{2k}$ with a function of
$H^\gamma_{1k}$, and the second term is a product of $H\gamma_{1k}$
with a function of $H^\gamma_{2k}$. Therefore we can write
\begin{align}
f^\gamma_{1k}=\exp(-a (H^\gamma_{1k})^2+c(f^\gamma_{1k}),
\end{align}
where $a$ is a positive real constant and $c(f^{\gamma}_{1k})$  is a
real constant. By the transform of switching the $ith$ and the $jth$
columns and rows ($i<j$), we can get that
\begin{align}
f^\gamma_{ij}=\exp(-a (H^\gamma_{ij})^2+c(f^\gamma_{ij})),
\end{align}
and since the sum of the first two terms is zero, we have
\begin{align}
f^\gamma_{ii}=\exp(-a H^\gamma_{ii}+b
H^\gamma_{ii}+c(f^\gamma_{ii})),
\end{align}
where $b$ and $c(f^\gamma_{ii})$ are constants.

By fixing the unitary matrix $U$ to be a diagonal matrix with the
entry at the $ith$ column and $ith$ row equal to $\vec{i}$, $
\vec{j}$, or $\vec{k}$, we  get for any $i<j$, $\gamma$,up to a
constant factor
\begin{align}
f^\gamma_{ij}=\exp(-a H^\gamma_{ij}),
\end{align}
and
\begin{align}
f^0_{ii} &=\exp(-a (H^0_{ii})^2+b H^0_{ii}),\nonumber\\
f^1_{ii} &=\exp(-a (H^1_{ii})^2),\nonumber\\
f^2_{ii}&=\exp(-a (H^2_{ii}),\nonumber\\
f^3_{ii}&=\exp(-a (H^3_{ii}).
\end{align}

Therefore we are able to write $P(H)$ by
\begin{align}
P(H)=\exp(- a \ tr HH^{+}+ b \ Re (tr H)+c),
\end{align}
where $c$ is a constant.

Conversely, if $P(H)=\exp(-Na  \ tr( HH^{+})+ Nb \ Re( tr H)+c)$, it is
straight forward to compute the correlation function of any two
components of two entries or one entry. Then the computation shows
that all the components of all the entries are statistically
independent.

\subsection{joint distribution of eigenvalues}
 The n point correlation function is defined by (\cite{M})
\begin{align}
R(\lambda_1,\lambda_2,\cdots,\lambda_n)=\frac{N!}{(N-n)!}\int P_N(\lambda_1,\cdots, \lambda_N)d\lambda_{n+1}\cdots\lambda_N.
\end{align}
As shown in \cite{M}, the normalized two point correlation fuction(in which the mean spacing at the origin is normalized to 1), is equal to $1-\frac{\sin(\pi r)}{\pi r}$.
Here we can prove for n point correlation fuction goes to a constant, if $|r_i-r_j|\rightarrow\infty$ for all $i,j$;
it goes to zero, if $|r_i-r_j|\rightarrow 0$ for all $i,j$. In fact, the normalized n point correlation function
\begin{align}
R_{norm}(x_1,x_2,\cdots,x_n)=\det(\frac{\pi}{(2N)^{1/2}}\sum_{k=0}^{N-1}\phi_k(x_i)\phi_k(x_j))_{1\leq i,j\leq n}
\end{align}
As $N\rightarrow\infty$, $\frac{\pi}{(2N)^{1/2}}\sum_{k=1}^{N-1}\phi_k(x_i)\phi_k(x_j)\rightarrow \frac{\sin(\pi r)}{\pi r}$, where $r=\frac{(2N)^{1/2}}{\pi}|x_i-x_j|$. Therefore we get the conclusion.

\subsection{how the $U(N)$ symmetry is breaking}

If the matrix elements are exactly gaussian, as we have proved,
then the gaussian complex ensemble has a $U(N)$ symmetry. However, physically the matrix elements
are not necessarily gaussian and this fact doesn't change the correlation function(Wigner's conjecture).
When all the eigenvalues are small in a ball of origin, $U(N)$ symmetry is approximated; when the eigenvalues
are far from the origin, the $U(N)$ symmetry is then broken.

\section{observables of  black holes}
In this model, gravitational phenomena can be understood as
following: we consider two gravitational systems in one Gaussian
unitary  ensemble, their spacetime distance observables correspond to
the norm of the substraction of eigenvalues of this ensemble. When the eigenvalues' norm are sufficiently large,
the joint distribution probability becomes smaller, then the classical gravitational phenomena arises;If the eigenvalues are restricted in a small vicinity of the origins, then because of the existence of the fermion determinant, there is a strong repulsion of the these eigenvalues from 0, which corresponds the black hole radiation.

  We are going to find the thermodynamics of black
hole physics in terms of random matrix theory. Classically, there
are four thermodynamical laws in classical black hole physics which
clearly indicate the thermodynamical nature of quantum gravity. A
static neutral black holes is determined by its mass $M$, angular
momentum $J$ and angular velocity $\Omega_H$ we have the generalized
Smarr formula
\begin{align}
M=2\Omega J_H+\frac{\kappa A}{4\pi}\label{4.1},
\end{align}
where $A$ is the area of the horizon.
In our present model, the symmetry is $U(N)$ which is maximized. Therefore the angular velocity is
zero. Since in the present random matrix model, all the matrix elements are statistically independent, then we
know the entropy
\begin{align}
S=C_1 N^2/\sqrt{a}.
\end{align}
where $C_1$ is the constant which equals to the entropy of each matrix element.
In classical black hole thermodynamcis,
\begin{align}
S=\frac{A}{4\hbar}=\frac{\pi r^2}{\hbar},
\end{align}
where $r$ is the radius of the black hole.

In information theory, for smooth probability distribution, there
are alternative definition of entropy which is called relative
entropy or Kullback¨CLeibler divergence. But here we use another
definition because it is approximate to the shannon entropy when $a$
is not large. The continuous version of shannon's entropy can be
negative. In fact,
\begin{align*}
I(P(H))&=-\int P(H)\ln P(H) dH\nonumber\\
       &=-\int \exp(-a tr(HH^{+})+b Re tr(H))  (\frac{a}{\pi})^{2N^2}\exp{(\frac{Nb^2}{4a})}
       \\&(-a tr(HH^{+})+bRe
       tr(H)+2N^2\ln a-2N^2 \ln\pi+\frac{Nb^2}{4a}) \prod_{\gamma=0}^3\prod_{1\leq i,j\leq N}dH_{ij}^{\gamma}\nonumber\\
       &=-\frac{1}{2^{4N^2}}-2N^2 \ln\frac{a}{\pi}
\end{align*}

In the following, we assume $a=1,b=0$.

In the present model,if we define the mass $M$ and $r^2$ proportional to
\begin{align}
 & N\langle z_1^2 \rangle_{Z_B(N)}\nonumber\\
  =& N C_{N2} \int_\mathbb{C} |z_1|^2  e^{-\frac{1}{2}\sum_{i=1}^N |z_i|^2}\prod_i
  |z_i-z_i^*|^2 \prod_{i<j}(|z_i-z_j|^2|z_i-z_j^*|^2 dz,  \label{mass}
\end{align}
where $C_{N2}$ is the normalization constant of GUE.
If we put the parameter $a$ in front of $\sum_{i=1}^N |z_i|^2$ in the joint probability distribution which is denoted
by $P(a)$, then its
 normalization constant is proportional to $a^{N^2+N}N!(2\pi)^N \prod_1^N \Gamma(2j)$(see chapter 17,\cite{M}), i.e.,
 \begin{align}
 \langle P(a) \rangle=const \  a^{N^2+N}N!(2\pi)^N \prod_1^N \Gamma(2j).
 \end{align}
Here the constant doesn't depend on $a$. Differentiating both sides, we get
\begin{align}
M\sim r^2 \sim N^2
\end{align}

\subsection{holographic principle}
Gerard 't Hooft proposed that in quantum gravity theoryl, the information of a gravitational system in spacetime is
totally determined by the area of the surface \cite{Hooft}. His conjecture is based on the entroply analysis of classical black hole theromodynamics.  From the present model, the  information or the degrees of the freedom of the underlying system is equal to the product of $N^2$ with a constant, which is proportional to the surface area $A=4\pi r^2$, where $r$ here is the schwartz radius in.

\subsection{the instability of Kerr black holes}
In this stationary black hole model, we have seen that the black
holes are neutral, which means a black hole will evaporate its
charge first than other matters. This is because if a charged black
hole is in a stationary state, it  has $SU(N)$ symmetry. However the
proof of the theorem indicates that the $SU(N)$ symmetry enhanced to
a $U(N)$ symmetry. Thus this black hole in fact is neutral.

\section{conclusion}
The issue of black hole is dealt with in the context of random matrix theory in this paper. Spacetime geometry and degrees of freedom are combined together by the virtue of Gaussian complex ensembles. This model also is also an asymptotically free theory. Classical black hole observables can be computed in this model and holographic principle is also preserved. However, how the four dimensional spacetime freedom arise is related to the $U(n)$ symmetry breaking process is  an important issue. We would like to expect and continue to work on this problem.

\end{document}